\documentclass[11pt]{article}
\usepackage{graphicx}

\newcommand{\BABARPubYear}    {06}

\newcommand{\BABARConfNumber} {006}
\newcommand{\SLACPubNumber} {12016}
\newcommand{\LANLNumber} {0000}

\input babarsym

\long\def\inst#1{\par\nobreak\kern 4pt\nobreak
    {\it #1}\par\vskip 10pt plus 3pt minus 3pt}

\newcommand{\Do}{D^0}

\newcommand{\Dstars}{{D}^{\ast +}_s}

\newcommand{\GeV}{\rm{GeV}}
\newcommand{\GeVc}{\rm{GeV/}c}
\newcommand{\GeVcd}{\rm{GeV/}c^2}

\newcommand{\MeVcd}{\rm{MeV/}c^2}

\newcommand{\ba}{\begin{array}}
\newcommand{\ea}{\end{array}}
\newcommand{\bc}{\begin{center}}
\newcommand{\ec}{\end{center}}
\newcommand{\beq}{\begin{eqnarray}}
\newcommand{\eeq}{\end{eqnarray}}
\newcommand{\bes}{\begin{eqnarray*}}
\newcommand{\ees}{\end{eqnarray*}}
\newcommand{\Zz}{\ifmmode {\rm Z} \else ${\rm Z } $ \fi}
\newcommand{\xxbar}{\ifmmode {\rm x\bar{x}} \else ${\rm x\bar{x}} $ \fi}
\newcommand{\rphi}{\ifmmode {\rm R\phi} \else ${\rm R\phi} $ \fi}

\begin{document}
{\pagestyle{empty}

\begin{flushleft}
\end{flushleft}

\begin{flushright}
\babar-CONF-\BABARPubYear/\BABARConfNumber \\
SLAC-PUB-\SLACPubNumber \\
hep-ex/\LANLNumber \\
July 2006 \\
\end{flushright}

\par\vskip 5cm

\begin{center}
\Large \bf 
Measurement of the \mbox{\boldmath $q^2$} Dependence of the Hadronic Form Factor in 
\mbox{\boldmath $\Do \rightarrow \Km e^+ \nu_e$} Decays. 
\end{center}
\bigskip

\begin{center}
\large The \babar\ Collaboration\\
\mbox{ }\\
\today
\end{center}
\bigskip \bigskip

\begin{center}
\large \bf Abstract
\end{center}

A preliminary measurement of the $q^2$ dependence of the 
$D^0 \rightarrow K^- e^+ \nu_e$ 
decay rate is presented.  This rate is proportional to the
hadronic form factor squared, specified by a single parameter. 
This is either
the mass in the simple pole ansatz 
$m_{\rm{pole}} = (1.854 \pm 0.016 \pm 0.020)~\rm{GeV}/c^2$ or the scale
in the modified pole ansatz 
$\alpha_{\rm{pole}}= 0.43 \pm0.03 \pm 0.04$. 
The first error refers to the statistical, the second to the systematic
uncertainty.

\vfill
\begin{center}

Submitted to the 33$^{\rm rd}$ International Conference on High-Energy Physics, ICHEP 06,\\
26 July---2 August 2006, Moscow, Russia.

\end{center}

\vspace{1.0cm}
\begin{center}
{\em Stanford Linear Accelerator Center, Stanford University, 
Stanford, CA 94309} \\ \vspace{0.1cm}\hrule\vspace{0.1cm}
Work supported in part by Department of Energy contract DE-AC03-76SF00515.
\end{center}

\newpage
} 

%
%
\begin{center}
\small

The \babar\ Collaboration,
\bigskip

%
{B.~Aubert,}
{R.~Barate,}
{M.~Bona,}
{D.~Boutigny,}
{F.~Couderc,}
{Y.~Karyotakis,}
{J.~P.~Lees,}
{V.~Poireau,}
{V.~Tisserand,}
{A.~Zghiche}
\inst{Laboratoire de Physique des Particules, IN2P3/CNRS et Universit\'e de Savoie,
 F-74941 Annecy-Le-Vieux, France }
% ...

\end{center}\newpage

\section{INTRODUCTION}
\label{sec:Introduction}
In exclusive semileptonic $B$ decays, the accuracy of the 
determination of the contributing Cabibbo, Kobayashi and Maskawa (CKM) 
matrix elements $\left | V_{cb} \right |$ and  $\left | V_{ub} \right |$
is limited by the precision of the corresponding
 hadronic form factors participating in these decays.
In $c$-hadron semileptonic decays, it can be assumed 
that $\left | V_{cs} \right |$ and  $\left | V_{cd} \right |$ are known
(considering for instance that the Cabibbo matrix is unitary). Measurements
of the hadronic form factors in such decays can thus be used to test 
predictions
on their absolute value and $q^2 ~(=(p_e+p_{\nu})^2)$ 
variation. In this expression, $p_e$ and $p_{\nu}$ are respectively
the 4-momenta of the positron and of the neutrino.

With the development of lattice QCD algorithms and the installation of
large computing facilities, accurate values for hadronic form factors are
expected from theory in the coming years. A comparison between these
values and corresponding measurements of a similar and possibly better accuracy
will validate these complex techniques and the uncertainties attached
to the remaining approximations.

In the present analysis the $q^2$ variation of the hadronic form  factor, 
in the decay $\Do \rightarrow \Km e^+ \nu_e$\footnote{Charge 
conjugate states are implied throughout this analysis.}, has been measured.
Neglecting the electron mass, there is a contribution from
a single form factor
and the differential decay rate is  a product of $q^2$-
and $\theta_e$- 
dependent expressions where
$\theta_e$ is the angle between the electron and the kaon
in the $e\nu_e$ rest frame. 
Integrating over the angular distribution,
the $q^2$ differential decay width reads: 
\beq
 \frac{d \Gamma}{d q^2} = \frac{G^2_F}{24 \pi^3} \left | V_{cs} \right |^2
p_K^3(q^2) \left |f_+(q^2) \right |^2.
\eeq

As this decay is induced by a vector current generated by
the $c$ and $\overline{s}$ quarks, the $q^2$ variation of 
the form factor $ f_+(q^2)$ is expected to be
governed by the $\Dstars$ pole. The following expressions have been
proposed \cite{ref:ks}:
\beq
\left |f_+(q^2) \right |= \frac{f_+(0)}{1-\frac{q^2}{m_{\rm{pole}}^2}},
\label{eq:pole}
\eeq
and \cite{ref:bk}
\beq
\left |f_+(q^2) \right |= \frac{f_+(0)}
{\left ( 1-\frac{q^2}{m_{D^*_s}^2}\right )
\left ( 1-\frac{\alpha_{\rm{pole}} q^2}{m_{D^*_s}^2}\right )}.
\label{eq:modpolemass}
\eeq

Equation (\ref{eq:pole}) is the ``pole mass'' and Equation (\ref{eq:modpolemass})
is the ``modified pole mass''.
Each distribution depends on a single parameter: $m_{\rm{pole}}$ and $\alpha_{\rm{pole}}$,
respectively. In Equation (\ref{eq:modpolemass})
$m_{D^*_s}=(2.1121\pm0.0007) ~\GeVcd)$ is the $D^*_s$ physical mass.
In lattice QCD computations, usually a ``lattice mass'' value
is used for $m_{D^*_s}$ but the computed value for $\alpha_{\rm{pole}}$
is expected to be directly comparable to the value extracted from the fit to  data
using expression (\ref{eq:modpolemass}).
A recent result from lattice QCD computations \cite{ref:lqcd}
gives $\alpha_{\rm{pole}}^{lattice}=0.50 \pm 0.04$.

Radiative effects in $\Do \rightarrow \Km e^+ \nu_e$ decays have been simulated
using the PHOTOS generator \cite{ref:photos} and present measurements assume
its validity.

\section{THE \babar\ DETECTOR AND DATASET}
\label{sec:babar}


Results given in this document have been obtained using \babar\ data taken
between February 2000 and June 2002,
corresponding to an integrated luminosity of $75$ fb$^{-1}$ and
which comprises the Run1 ($18.6$ fb$^{-1}$) and Run2 ($56.4$ fb$^{-1}$)
data taking periods. 
In addition, Monte Carlo (MC) simulation samples of 
charm, $b$-hadrons and light quark pairs, equivalent to 
about $1.3$ times the data statistics have been used. 

The \babar\ detector is described elsewhere~\cite{ref:babar}. Its most
important capabilities for this analysis are charged-particle tracking and momentum measurement,
charged $\pi/K$ separation, electron identification, and photon reconstruction. 
Charged particle tracking is provided
by a five-layer silicon vertex tracker and a 40-layer drift chamber. The 
Detector of Internally Reflected Cherenkov light (DIRC), a Cherenkov ring-imaging 
particle-identification system, is used to separate charged kaons and pions. Electrons are identified
using the electromagnetic calorimeter, which comprises 6580  thallium-doped CsI crystals. These
systems are mounted inside a 1.5 T solenoidal superconducting magnet.

\section{ANALYSIS METHOD}
\label{sec:Analysis}

%
This analysis is based on the reconstruction of $\Dstarp$ mesons produced
in continuum $c\overline{c}$ events and in which 
$\Dstarp\rightarrow D^0 \pi^+$ and 
the $D^0$ decays semileptonically. 

\subsection{Candidate selection and background rejection}
Electron candidates are selected with a momentum larger than
0.5 $\GeVc$ in the laboratory and, also, in the center of mass frame (cm).
Muons are not used in this analysis as they have less purity than electrons
and as we are not limited by statistics.
The direction of the event thrust axis is taken in the interval
$|\cos(\theta_{\rm{thrust}})|<0.6$ to minimize the loss of particles in 
regions close to the beam axis and to ensure a good 
total energy reconstruction. 

Two variables, R2 (the ratio between the second and zeroth
order Fox-Wolfram moments \cite{ref:r2}) and the total charged and neutral 
multiplicity,
 have been used to reduce the contribution from $B$ events
based on the fact that the latter are more spherical.
These variables have been combined linearly in a Fisher discriminant
\cite{ref:fisher} 
on which a selection requirement retains $71\%$ of signal events and $10\%$ of
the $B$ background. 

Charged and neutral particles are boosted to the center of mass system 
and the event thrust
axis is determined. A plane perpendicular to 
the thrust axis, and containing the beam interaction point,
is used to define two event's hemispheres.
Each hemisphere is considered in turn to search for
a candidate having an electron($\pm$), a kaon($\mp$) and a 
pion($\pm$) reconstructed in that hemisphere and with the relative charges
as given within parentheses. A candidate triplet will be called also a right 
sign combination (RS), other different relative charge assignments 
will be called wrong sign (WS). 
 
To evaluate the neutrino ($\nu_e$) momentum, 
two constrained fits are used. In the first fit 
the $(e^+ \Km \nu_e)$ invariant mass is constrained to be equal to the $\Do$ mass 
whereas, in the second fit, the $(e^+ \Km \nu_e \pi^+)$ invariant mass 
is constrained to be equal to the $\Dstarp$ mass. 
In these fits, estimates of the $\Do$ direction and of the neutrino energy are 
included from measurements obtained from all particles registered in the event.
The $\Do$ direction estimate is taken as the direction of the vector opposite 
to
the momentum sum of all reconstructed particles but the kaon and the
electron. The neutrino energy is evaluated by subtracting from the hemisphere
energy, the energy of reconstructed particles contained in that hemisphere.
The energy of each hemisphere has been evaluated by considering that the
total center of mass energy is distributed in two objects of mass corresponding
to the measured hemisphere masses. The hemisphere mass is
the mass of the system corresponding to the sum of the 4-vectors for
particles contained in that hemisphere. 
Detector performance in the reconstruction of the $\Do$ direction and missing
energy have been measured using events in which the $\Do \rightarrow \Km \pi^+$
and used in the mass constrained fits.

\begin{figure}[!htb]
\begin{center}
\includegraphics[height=9cm]{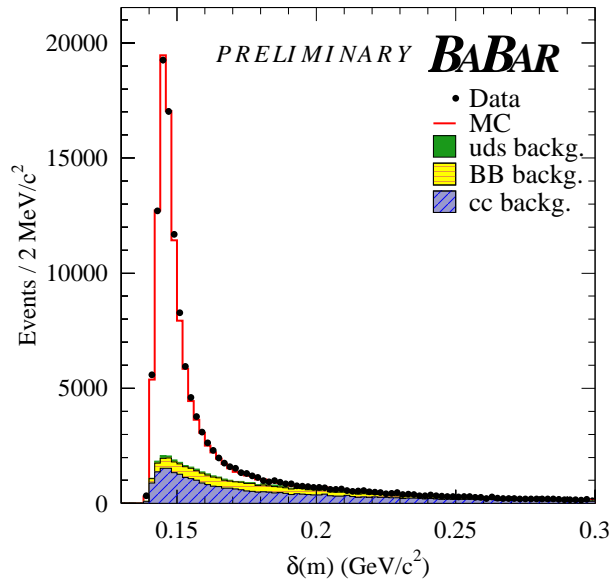}
\caption{ $\delta(m)$ distributions from data and simulated events. 
The signal and the different background components are indicated.
MC events have been normalised to the data luminosity according to the 
different cross sections. The light quark component is very  
small and hardly visible.}
\label{fig:deltamback}
\end{center}
\end{figure}

After performing the first fit, the events with a $\chi^2$ probability larger than 
$10^{-3}$ are kept.
The $\Do$ 4-momentum is then obtained and the mass difference
$\delta(m) = m(\Do \pi^+)-m(\Do)$ is evaluated and is shown in Figure 
\ref{fig:deltamback}.

Other particles, present in the hemisphere, which are not decay products
of the $\Dstarp$ candidate are named ``spectator'' tracks. They originate from
the beam interaction point and are emitted during hadronization
of the created $c$ and $\overline{c}$ quarks. The 
``leading'' track is the spectator track having the largest momentum. 
Information from the spectator system is used, in addition to the one provided
by variables related to the $\Do$ production and decay, to reduce the 
contribution from the combinatorial background. As charm hadrons take a large
fraction of the charm quark energy, charm decay products
have higher average energies than spectator particles. 

The following variables have been used:

\begin{itemize}
\item the $\Do$ momentum after the first mass constrained fit;
\item the spectator system mass, which has lower values for signal events;
\item the direction of the spectator system momentum relative to the thrust axis;
\item the momentum of the leading spectator track;
\item the direction of the leading spectator track relative to the $\Do$ direction;
\item the direction of the leading spectator track relative to the thrust axis;
\item the direction of the lepton relative to the kaon direction, in the 
virtual W rest frame;
\item the charged lepton momentum in the cm frame.
\end{itemize}
The first six variables depend on the properties of
charm quark hadronization whereas the last two are related to decay
characteristics of the signal.
These variables have been linearly combined into a Fisher discriminant 
variable  
and events have been kept for values above 0, retaining  $82\%$
of signal events and rejecting $52\%$ of background candidates.

The remaining background from $c \overline{c}$ events can be divided into peaking and 
non-peaking categories. Peaking events in  
$\delta(m)$ have a real $\Dstarp$ in which the slow $\pi^+$ is
included in the candidate track combination. 
The non-peaking background corresponds to candidates without a 
charged  $\Dstarp$ slow pion. 
Using simulated events, these components are displayed
in Figure \ref{fig:deltamback}. The MC values have been rescaled to the 
data luminosity, using the cross sections 
of the different components (1.3 nb for $c\overline{c}$, 0.525
nb for $\Bp\Bm$ and $B^0 \bar{B}^0$, 2.09 nb for light $uds$ quark events).

To study the $q^2$ distribution, events with $\delta(m)$ below 0.16 $\GeVcd$
and with a $\chi^2$ probability, of the second mass constrained fit greater than 1$\%$ have been
selected.

\subsection{$q^2$ measurement}
The $\Do \rightarrow \Km e^+ \nu_e$ is a 3-body decay and its dynamics
depends on two variables. In semileptonic decays these variables are usually
taken as $q^2$ and $\cos(\theta_e)$.
Throughout the present analysis, $q^2$ has been evaluated 
using $q^2=\left ( p_D-p_K \right )^2$ 
$\left (=\left ( p_e+p_{\nu_e} \right )^2 \right )$ 
where $p_D$ and $p_K$ are the four-momenta of the $D$ and  $K$ meson, respectively.  
The differential decay rate versus these two variables factorises
in two parts depending, respectively, on $q^2$ and $\theta_e$. 
The variation of $d\Gamma / d\cos(\theta_e)$ is fixed by angular momentum and 
helicity conservation and has a $\sin^2{\theta_e}$ dependence.
The aim of the present analysis is to measure the $q^2$ dependence of the form
factor whose shape depends on the decay dynamics.

\begin{figure}[!htb]
\begin{center}
\includegraphics[height=9cm]{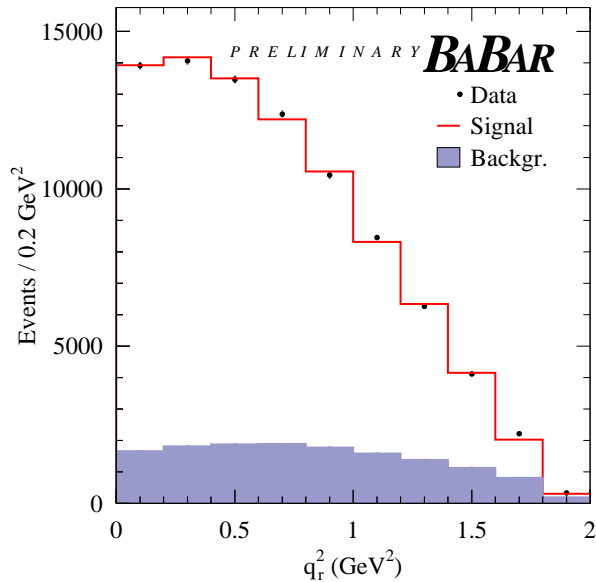}
\caption{ Distributions of $q^2_r$. Points with error bars are data.
The distribution from background events is given by the blue histogram.
The fitted signal component is overlayed.}
\label{fig:q2rall}
\end{center}
\end{figure}

After having applied the $\Do$ and $\Dstarp$ mass constrained fits, 
the values of the variables 
$q^2$ and $\cos(\theta_e)$  are obtained
and will be noted as $q^2_r$ and $\cos(\theta_e)_r$, respectively.
Background contributions in these distributions have been evaluated from
corresponding distributions obtained in simulated event samples,
normalized to the same integrated luminosity analyzed in data. 
The measured $q^2_r$ distribution is shown in Figure \ref{fig:q2rall}.
The expected background level has been indicated and the fitted signal
component, as obtained in the following, has been added.
The $q^2_r$ distribution for signal events
is obtained by subtracting the estimated background distribution from
the total measured distribution. 
Using simulated events the reconstruction accuracy on $q^2$ has been 
obtained by comparing reconstructed ($q^2_r$) and true simulated ($q^2_s$) values.
\begin{figure}[!htb]
\begin{center}
\includegraphics[height=9cm]{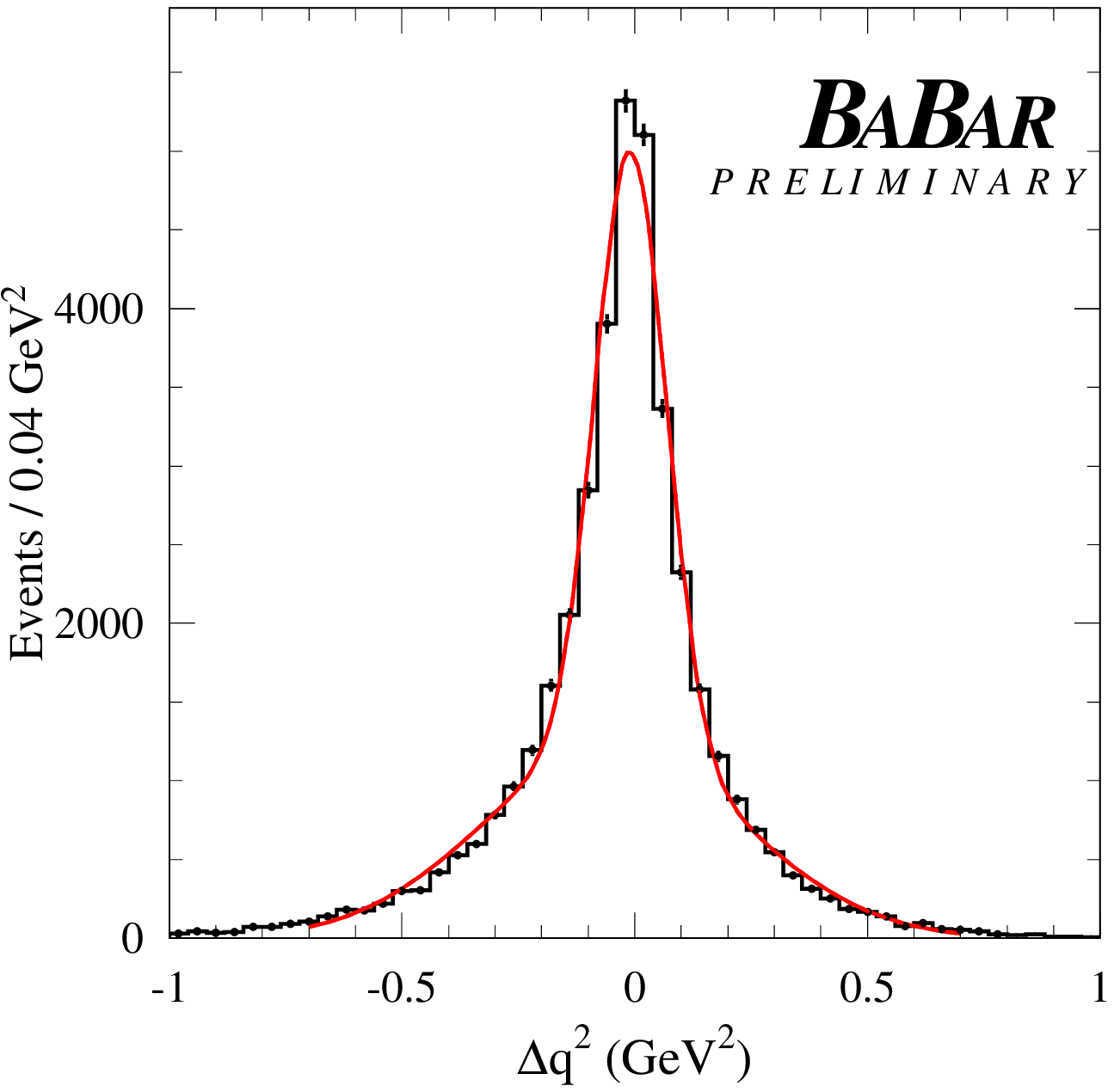}
\caption{ Distribution of the difference between the real and the
reconstructed value of the $q^2$ variable ($\Delta q^2 = q^2_r - q^2_s$)
as obtained from
simulated events. The fitted curve is the sum of two Gaussian distributions.}
\label{fig:q2res}
\end{center}
\end{figure}
Figure \ref{fig:q2res} shows a double Gaussian fit to the $q^2$ resolution distribution. 
We obtain widths $\sigma_1=0.077 ~\GeV^2$ and $\sigma_2=0.276 ~\GeV^2$
for the two Gaussians, which are of similar area.

\begin{figure}[!htb]
\begin{center}
\includegraphics[height=9cm]{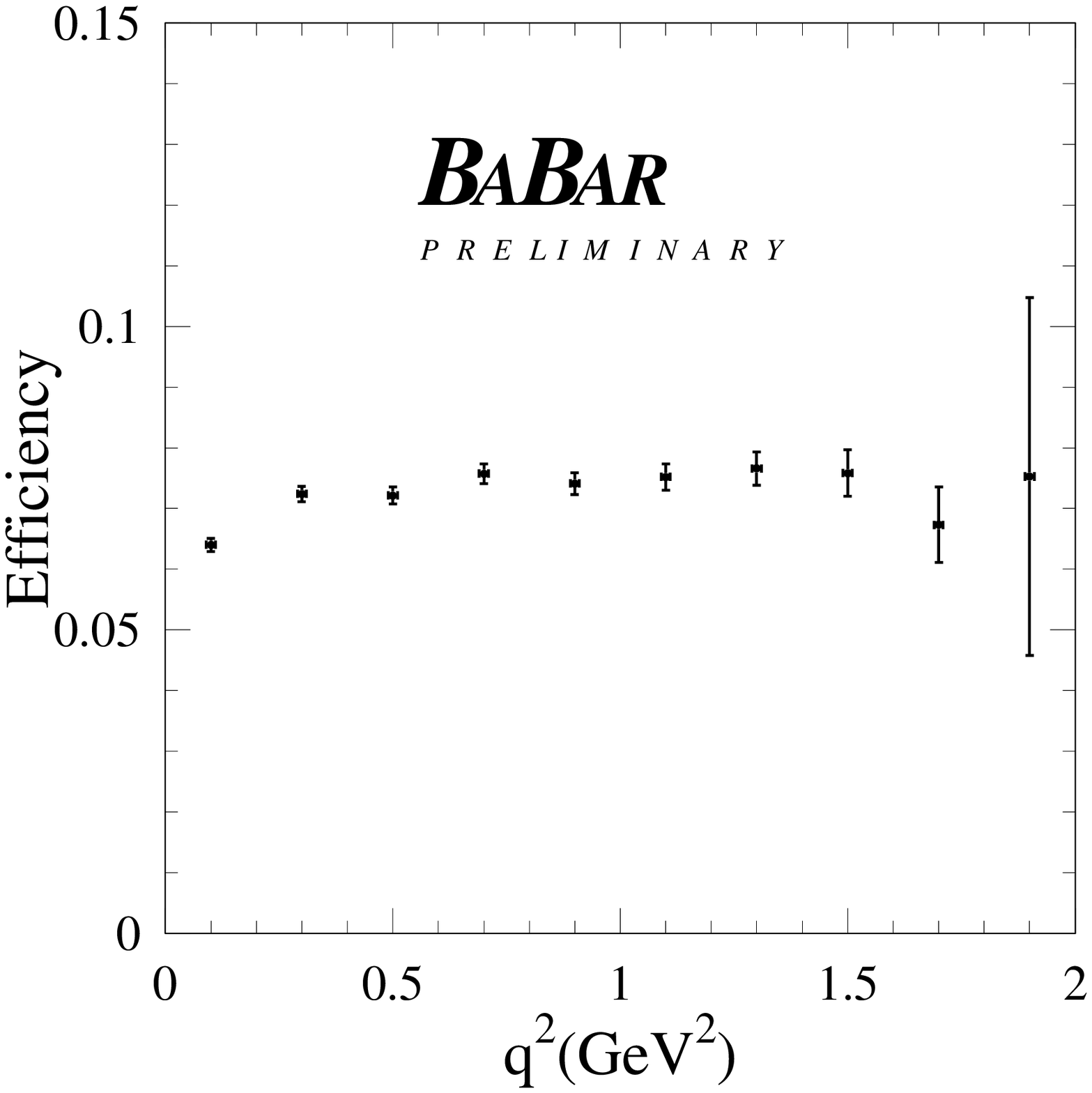}
\caption{ $q^2_s$ dependent efficiency after all selection requirements, 
from (Run1) simulated signal events.}
\label{fig:effi}
\end{center}
\end{figure}

The variation of the signal efficiency as a function of $q^2_s$
is shown in Figure \ref{fig:effi}.

\subsection{Unfolding procedure}
To obtain the unfolded $q^2$ distribution for signal events, 
corrected for resolution 
and acceptance effects,
the Singular Value Decomposition (SVD) 
\cite{ref:svd} of the resolution matrix
has been used in conjunction with a regularization which minimizes
the curvature of the correction distribution.
A SVD of a real $m\times n$ matrix $A$ is its factorization of the form:
\beq
A~=~U~S~V^T,\nonumber
\eeq
where $U$ is an $m\times m$ orthogonal matrix, $V$ is an $n\times n$ orthogonal matrix,
while $S$ is an $m\times n$ diagonal matrix with non-negative elements:
\beq
S_{ii}=s_i \geq 0. \nonumber
\eeq
The quantities $s_i$ are called Singular Values (SV) of the matrix $A$.
If the matrix of a linear system is known with some level of uncertainty,
and some SV of the matrix are significantly smaller than others, the system may be 
difficult to solve even if the matrix has full rank, and SVD suggests a
method of treating such problems. We will assume that the SV values $s_i$
form a non-increasing sequence.

The method uses binned distributions. It needs, as input, a 2-d array
which indicates how events generated in a bin in $q^2_s$ are distributed
over several bins in $q^2_r$. To be able to correct
for acceptance effects one needs also the initial
distribution in $q^2_s$, as given by the generator.
The number of events estimated in each bin of the $q^2_r$
variable $\left ( n^{i,estimated}_{q^2_r} \right )$ can
be expressed as:
\beq
\sum_{j=1}^{m} A_{ij} w^j = n^{i,estimated}_{q^2_r}
\nonumber
\eeq 
where $w^j$ is an unknown deviation of the number of simulated events
in bin $j$ of the $q^2_s$ variable, from the initial Monte Carlo
value. $A_{ij}$ is the actual number of events which were generated in
bin $j$ and ended up in bin $i$. The elements of the vector $w^j$
are determined by minimizing a $\chi^2$ expression obtained
by comparing estimated and measured number of events in each 
$q^2_r$ bin:
\beq 
\sum_{i=1}^{n}{ \left ( \frac{\sum_{j=1}^{m}{ A_{ij} w^j -n^i_{q^2_r}}}{\sigma_{n^i_{q^2_r}}} \right )^2} = min,
\nonumber
\eeq
where $\sigma_{n^i_{q^2_r}}$ is the corresponding uncertainty on the measured number
of events in bin $i$. The minimization leads to the system: 
\beq
\tilde{A} w = \tilde{n}_{q^2_r}
\label{eq:svdmin}
\eeq
where $\tilde{A}_{ij}=A_{ij}/\sigma_{n^i_{q^2_r}}$ and
$\tilde{n}_{q^2_r}= n^i_{q^2_r}/\sigma_{n^i_{q^2_r}}$.

The exact solution of Equation (\ref{eq:svdmin})
will again most certainly lead to  rapidly oscillating distribution. This spurious
oscillatory component should be suppressed and this can be achieved by adding a 
regularization term to the expression to be minimized:
\beq
\left ( \tilde{A}w - \tilde{n} \right )^T\left ( \tilde{A}w - \tilde{n} \right )
+\tau \left ( Cw\right )^T Cw ~=~ min.
\nonumber
\eeq
Here $C$ is a matrix which suppress solutions $w$ having large curvatures and $\tau$
determines the relative weight of this condition. The curvature of the discrete
distribution $w_j$ is defined as the sum of the squares of its second derivatives:
\beq
\sum_i\left [ \left( w_{i+1}-w_i \right ) - \left( w_{i}-w_{i-1} \right ) \right ]^2.
\nonumber
\eeq

The unfolded $q^2$ distribution is given by the bin-by-bin product
of $w$ and of the initial Monte Carlo distribution. 
This approach provides the full covariance matrix for the
bin contents of the unfolded distribution. Its inverse should be used
for any $\chi^2$ calculation involving the unfolded distribution.

For a square response matrix, Singular Values (SV) correspond essentially to eigenvalues
of this matrix. 
The number of significant SV has been obtained by looking at
their distribution in experiments generated using 
a toy simulation
of the experimental conditions, and keeping those which are above
a plateau \cite{ref:svd}. According to this 
procedure, four values can be retained
in the present analysis. 

Unfolded $q^2$ distributions, obtained by dividing the total 
sample into four similar data sets (each corresponding
to about 21k events) have been analyzed.
These distributions, and the total resulting sample, have been fitted 
using the pole and the 
modified pole models. Corresponding values for the parameters are 
given in Table \ref{tab:statres}.

It has to be noted that obtained values for $m_{\rm{pole}}$ and $\alpha_{\rm{pole}}$
are independent of a choice for a number of SV. In addition
selecting a fixed number of SV results in introducing biases
on the values evaluated in each bin for the unfolded distribution. Thus, at present,
when providing the $q^2$ dependence of the hadronic form factor and the 
corresponding error matrix we have kept all SV 
(see also Section \ref{sec:systsvd}).

\begin{table}[htb]
  \caption {Fitted values for the parameters corresponding respectively
to a pole mass and a modified pole mass model for the form factor, 
obtained in different samples of similar statistics and
for the total. The last column
indicates the fraction of background events.}
\begin{center}
  \begin{tabular}{c|c c c}
    \hline
\hline
Data sample & $m_{\rm{pole}}$ $(\GeVcd)$ & $\alpha_{\rm{pole}}$ &bckg. fraction\\
\hline
Run1 & $1.808 \pm0.027$ & $0.54 \pm0.06$ & $16.3 \%$  \\
Run2-1 & $1.889 \pm0.033$ & $0.37 \pm0.07$ & $16.6 \%$  \\
Run2-2 & $1.847 \pm0.032$ & $0.43 \pm0.07$ & $16.4 \%$  \\
Run2-3 & $1.864 \pm0.033$ & $0.41 \pm0.07$ & $16.7 \%$  \\
\hline
All & $1.854 \pm0.016$ & $0.43 \pm0.03$ & $16.5 \%$  \\
\hline
\hline
  \end{tabular}
\end{center}
  \label{tab:statres}
\end{table}

\section{SYSTEMATIC STUDIES}
\label{sec:Systematics}

Systematic uncertainties originate from non-perfect simulation
of the charm fragmentation process and 
of the detector response, uncertainties in the control
of the background level and composition, and the unfolding
procedure. 
Differences between data and Monte Carlo in quantities used in 
the analysis may result in biases that need to be corrected.

These effects have been evaluated by redoing the unfolding procedure
after having modified the conditions that correspond to
a given parameter entering in the systematics and evaluating the variation
on the value of the fitted
$m_{\rm{pole}}$ and $\alpha_{\rm{pole}}$ parameters.

For some of these studies dedicated event samples have been used
in which a $\Do$ is reconstructed in the $\Km \pi^+$ or $\Km \pi^+ \pi^0$
decay channel.

\subsection{Systematics related to $c$-quark hadronization}

The signal selection is based on variables related to $c$-quark
fragmentation and decay properties  of signal events. As we
measured differences between the hadronization properties of 
$c$-quarks events in real and simulated events, a $q^2$-dependent
difference in efficiency between these two samples is expected.
To have agreement between the distributions measured in real
and simulated events, for the different variables entering
into the Fisher analysis a weighting procedure has been used.
These weights have been obtained using events with a reconstructed
$\Do$ decaying into $\Km \pi^+$ or $\Km \pi^+ \pi^0$.
The data-MC agreement
of the measured distributions 
indicates
that uncertainties related to the tuning are below 5 $\MeVcd$ on $m_{\rm{pole}}$.
These corrections correspond to a 16 $\MeVcd$ displacement on $m_{\rm{pole}}$.

\subsection{Systematics related to event selection and analysis algorithm}
Another study has been done by analyzing $\Do \rightarrow \Km \pi^+ \pi^0$
events as if they were $\Km e^+ \nu$ events.
The two photons from the $\pi^0$ are removed from the particle lists and 
events are reconstructed using the algorithm applied to the semileptonic
$\Do$ decay channel. The ``missing'' $\pi^0$ and the charged pion 
play, respectively, the role of the neutrino and of the electron. 
In the two mass constrained fits, to preserve the correct kinematic
limits, it is necessary to consider that the ``fake'' neutrino has the
$\pi^0$ mass and that the ``fake'' electron has the $\pi^+$ mass.
With these events one can measure
possible $q^2$-dependent effects related to some of the selection
criteria 
and  the 
accuracy of the measurement of $q^2$.

In these studies the ``exact'' value
of $q^2$ is defined as $q^2=(p_{\pi^+}+p_{\pi^0})^2$.
Results obtained using real and simulated events are
compared to identify possible differences.
Fractions of real and simulated events are
selected using the 
Fisher discriminant variables
defined to reject background events, the $\chi^2$
probability from the two fits, the  $\delta(m)$ selection, and
the requirement for having at least one spectator particle. 
In this selection, the Fisher variable used against the $c\overline{c}$ background 
does not include $\cos(\theta_e)$ and $p_e$. This is because the angular 
distribution in $\Km \pi^+ \pi^0$ events is close to 
$\cos^2(\theta_{\pi})$.

There is no evidence for a statistically significant different behavior 
in data as compared with the simulation. 
The statistical accuracy of this comparison, 
based at present on the Run1 data sample,  corresponds to an uncertainty
on the pole mass of $\pm 15 ~ \MeVcd$.

\subsection{Detector related systematics}

To search for differences between real and simulated events, distributions 
of $\Delta q^2=q^2_r-q^2_s$, obtained by selecting 
$\Do \rightarrow \Km \pi^+ \pi^0$ 
events in a given
bin of $q^2_s$  have been compared. One notes that these distributions
 are systematically slightly narrower for simulated events.
We have evaluated this effect to be lower than 5$\%$ and have redone
the exercise after adding a smearing on $q^2_r$, in simulated events.
We obtain a +4 $\MeVcd$ increase on the fitted pole mass and
assign this value as a systematic error.

Effects from a momentum dependent difference between real 
and simulated events on
the charged lepton efficiency reconstruction and on the kaon
identification have been evaluated.
Such differences have been 
measured for electrons and kaons using dedicated data samples.
When applied, the corresponding corrections induce an increase of the 
fitted pole 
mass 
by +2 $\MeVcd$ and +4 $\MeVcd$, for electrons and kaons respectively.
A relative uncertainty of 30$\%$ has been assumed on this correction.

\subsection{Background related systematics}

The background under the $\Dstarp$ signal has two components which have
respectively a peaking and a non-peaking behavior.

The peaking background arises from events with a real $\Dstarp$
in which the slow $\pi^+$ is included in the candidate track
combination. Its main components, as expected from
the simulation, comprise events with real or fake $\Km$ or positron.
They mainly correspond to $\Do \rightarrow \Km \pi^0 e^+ \nu_e$ and 
$\Do \rightarrow \Km \pi^+ \pi^0$ events where, for the latter, the $\pi^0$
has a Dalitz decay or a converted photon.

\begin{figure}[!htb]
\begin{center}
\includegraphics[height=9cm]{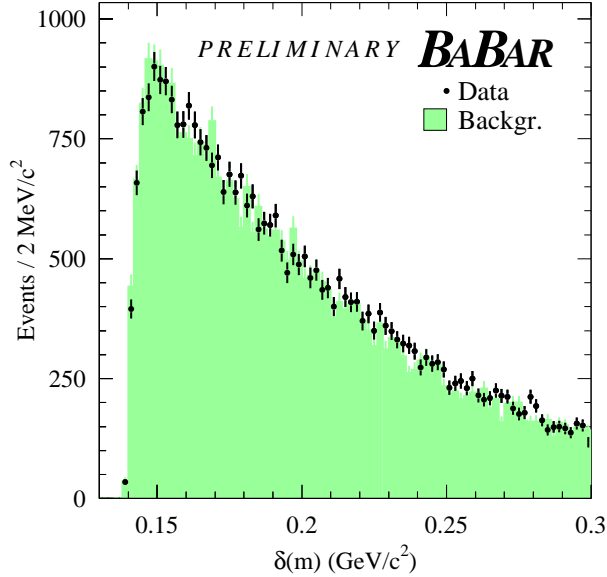}
\caption{ $\delta(m)$ distributions from data (points with error bars)
and simulated events (shaded histogram) in WS $\Km e^+ \pi^- $ events.} 
\label{fig:deltambackws}
\end{center}
\end{figure}

The non-peaking background originates from non-$c\overline{c}$ events
and from continuum charm events in which the $\pi$ candidate does not
come from a decaying $\Dstarp$. RS combinations, for
$\delta(m)>0.18~\GeVcd$ and WS combinations can be used to 
compare data and simulated event rates. 
It has been verified that $\delta(m)$ distributions were in agreement
in the tail of RS events and reproduce well the level of WS combinations
(see Figures \ref{fig:deltamback} and \ref{fig:deltambackws}). 

\begin{table}[!htbp]
  \caption {Systematics corresponding to uncertainties on
the level of the different background sources.}
\begin{center}
  \begin{tabular}{c|c c c }
    \hline
\hline
Source & variation & $\delta \left ( m_{\rm{pole}}\right )$  & $\delta \left ( \alpha_{\rm{pole}}\right )$  \\
       & $(\%)$ & $(\MeVcd)$ & \\
\hline
$\Do \rightarrow \Km \pi^0 e^+ \nu_e$  & $\pm9$ & $\pm4.4$ &$\mp0.0087$  \\
other peaking  & $\pm10$ & $\pm1.2$ & $\mp0.0023$  \\
$c\overline{c}$ non-peaking  & $\pm10$ & $\pm3.0$ & $\mp0.0062$  \\
$\Bp \Bm$  & $\pm10$ & $\pm2.6$ &$\mp0.0050$  \\
$B^0 \bar{B}^0$   & $\pm10$ & $\pm4.3$ & $\mp0.0084$  \\
$uds$   & $\pm10$ & $\pm0.5$ & $\mp0.0010$  \\
    \hline
total  &     & $\pm7.4$ & $\pm0.0147$ \\
    \hline
\hline
  \end{tabular}
\end{center}
  \label{tab:syst1}
\end{table}

As outlined in Table \ref{tab:syst1},
each background component has been varied  by $\pm10\%$
( a value which is larger than observed differences in the analyzed distributions),
apart for the $\Do \rightarrow \Km \pi^0 e^+ \nu_e$ component for which we 
use $\pm9\%$, corresponding to the combination of the PDG uncertainty
with a recent measurement of this decay channel by the CLEO-c experiment
\cite{ref:cleoc}.

\subsection{Control of the statistical accuracy and of systematics in the 
SVD approach}
\label{sec:systsvd}
For a fixed number of singular values, we verify that the statistical 
precision obtained for each binned unfolded value is reasonable and that 
biases generated by having removed information  are under control.

This has been achieved by fitting the same 
model for the form factor, directly
on the measured distribution, 
after having included resolution effects, 
and then by comparing the uncertainty obtained
on the extracted parameter with the one determined by fitting the unfolded 
distribution using that model. These studies have been complemented
by toy simulations.

One observes that the error obtained from a fit of the unfolded
distribution is underestimated by a factor of $\sim 1.3$, 
which depends on the statistics of 
simulated events.
Pull distributions indicate also that unfolded values in each bin are biased.
The importance of the biases decreases as the number of kept singular values increases. When
the number of SV is equal to the number of bins no biases are present.
When the number of SV is smaller,
these effects are corrected using pull results. It has been verified 
that these
corrections, obtained for a given value of the pole
mass, do not bias results expected in experiments generated
with a different value for $m_{\rm{pole}}$. Corrections defined with $m_{\rm{pole}}=1.8~\GeVcd$
induce a shift below 4 $\MeVcd$ when applied on experiments
generated with $m_{\rm{pole}}=1.7~{\rm or}~1.9~\GeVcd$.

\subsection{Systematics summary}
The systematic uncertainties evaluated 
on the determination of $m_{\rm{pole}}$ and $\alpha_{\rm{pole}}$ are summarized
in Table \ref{tab:systall}.

\begin{table}[htbp]
  \caption {Summary of corrections and systematics on the fitted 
parameters. The sign of the correction is given by the sign
of the difference between the fitted values after and before
applying the correction.}
\begin{center}
  \begin{tabular}{c|c c }
    \hline
\hline
Source & $\delta \left ( m_{\rm{pole}}\right )$ & $\delta \alpha_{\rm{pole}}$\\
       &  $(\MeVcd)$ & \\
\hline
$c$-hadronization tuning & $-16\pm 5$ & $+0.03\pm 0.01$  \\
reconstruction algorithm & $\pm 15$ & $\pm 0.03$  \\
resolution on $q^2$  & $+4\pm 4$ & $-0.01\pm 0.01$  \\
particle ID  & $+6\pm 2$ & $-0.02\pm 0.01$  \\
background control  & $\pm 7$ & $\pm 0.02$  \\
unfolding method &  $\pm 5$ & $\pm 0.01$  \\
    \hline
total  & $-6\pm 20$    & $ +0.01\pm 0.04$ \\
    \hline
\hline
  \end{tabular}
\end{center}
  \label{tab:systall}
\end{table}

The systematic error matrix for the ten unfolded values has been also
computed by considering, in turn, each source of uncertainty and by measuring 
the
variation, $\delta_i$, of the corresponding unfolded value in each bin
($i$). The elements of the error matrix are the sum, over all sources of
systematics, for the quantities $\delta_i \cdot \delta_j$. 
The total error matrix has been evaluated as the sum of the matrices
corresponding respectively to statistical and systematic uncertainties.

\section{RESULTS}
\label{sec:Physics}


In the present measurement the fit to a model has been done by
comparing the number of events measured in a given bin of $q^2$
with the expectation from the exact analytic integration
of the expression $p_K^3(q^2) \left |f_+(q^2) \right |^2$
over the bin range. The normalisation is left free to float and the 
weight matrix evaluated from SVD is used.  
From the measured integrated decay spectrum in each bin, the value
of the form factor $|f_+(q^2)|$ has been evaluated  at the bin
center. By convention it has been assumed that $|f_+(0)|=1$.

\begin{figure}[!htb]
\begin{center}
\includegraphics[height=9cm]{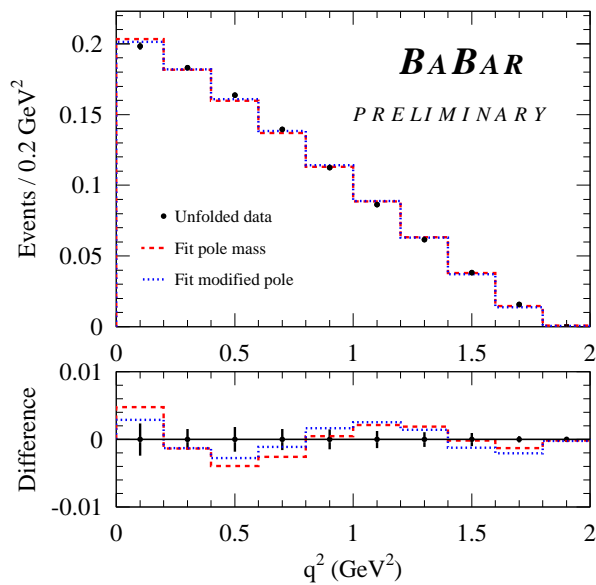}
\caption{ Comparison between the normalised unfolded $q^2$ distribution 
of the form factor obtained in the present analysis, keeping four SV,
and those
corresponding to the two fitted models. Lower plots give the difference
between measured and fitted distributions. Vertical lines correspond
to statistical uncertainties.}
\label{fig:q2datasim}
\end{center}
\end{figure}

The unfolded $q^2$ distribution, for signal events 
and keeping four SV, is given 
in Figure \ref{fig:q2datasim} where it has been compared 
with the fitted distributions obtained using the pole and
the modified pole ansatz.

\begin{figure}[!htb]
\begin{center}
\includegraphics[height=9cm]{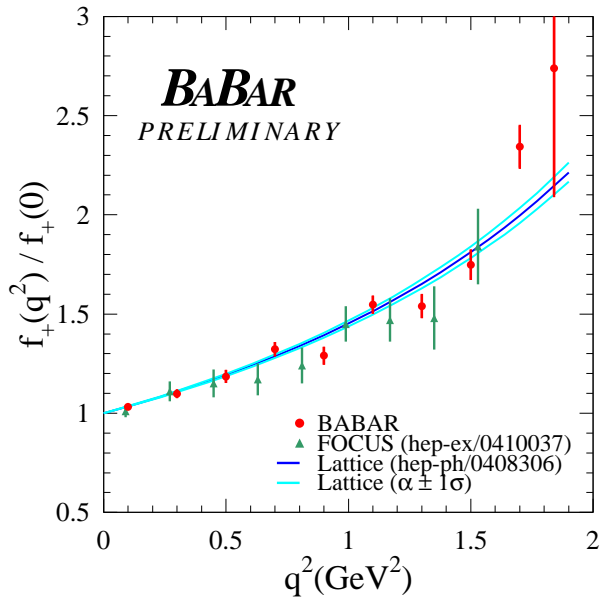}
\caption{ Comparison between the measured variation of $<|f_+(q^2)|>$
obtained in the present analysis and in the FOCUS experiment.
The band corresponds
to lattice QCD \cite{ref:lqcd} expectations.}
\label{fig:fq2}
\end{center}
\end{figure}

Results for the values of $m_{\rm{pole}}$ and $\alpha_{\rm{pole}}$
are independent of the choice of a given number of kept
SV. 
The number of SV corresponds to the number of independent combinations
of the ten bins used in the analysis. To minimize correlations between
bin's content and to compare with the $q^2$
distributions obtained in other experiments the distribution obtained 
using ten SV has been provided. 
The content of each bin is given in Table \ref{tab:errmeas}. 
The corresponding statistical and total error matrices are 
given also in that table and the $q^2$ variation of the hadronic
form factor is displayed in Figure \ref{fig:fq2}.

\begin{table}[htbp]
  \caption {Unfolded distribution using ten SV. 
The second line of this table gives the integrated values of the 
differential decay 
branching fraction
over 0.2$(\GeV^2)$ intervals (quoted on the first line). 
The total distribution
has been normalised to unity for $q^2$ varying from 0 to 2 $(\GeV^2)$.
The statistical error matrix corresponding to the above measurements
is then provided. 
On the diagonal is given the uncertainty
on each measured value. Off-diagonal terms correspond to the 
correlation coefficients. Including systematics, the total
error matrix is then provided.}
\begin{center}
\scriptsize{
  \begin{tabular}{|c|c|c|c|c|c|c|c|c|c|c|}
    \hline
$q^2$ bin&[0, 0.2] &[0.2, 0.4] &[0.4, 0.6] &[0.6, 0.8] &[0.8, 1.0] &[1.0, 1.2] &[1.2, 1.4] &[1.4, 1.6] &[1.6, 1.8] &[1.8, 2.0] \\
\hline
fraction &0.1999   &   0.1791   &    0.1606   &    0.1489    &   0.1007 &
 0.0962  & 0.0570  & 0.0363  & 0.0201  & 0.0012\\

\hline
stat. &0.0040 & -.63 & .24& -.094 & .034 & -.0065 &-.0022 &0.0006 & -.0025 &.0002 \\
error&& 0.0072 & -.66 & .27 & -.10 & .025 & .0015 & -.0003 & .0008 &-.0001 \\
and && & 0.0090 & -.69 & .29 & -.081 & .011 & -.0074 & .0022 & .0004 \\
correl.& & & & 0.0081 & -.67 & .24 & -.060 & .025 & -.0063 & .0019 \\
& & & & & 0.0071 & -.64 & .23 & -.083 & .021 & -.0066 \\
& & & & & & 0.0056 & -.64 & .25 &-.064 &-.0035 \\
& & & & & & & 0.0044 & -.60 & .15 & .022 \\
& & & & & & & & 0.0032 & -.44 & -.050 \\
& & & & & & & & & 0.0018 & -.059 \\
& & & & & & & & & & 0.00054 \\
    \hline

\hline
total &0.0045 & -.48 & .23 & -.097 & -.0084 & -.073 &-.079 &-.082 & -.12 &-.044 \\
error&& 0.0073 & -.65 & .27 & -.11 & .0038 & -.023 & -.027 & -.039 & -.016 \\
and && & 0.0090 & -.69 & .28 & -.086 & .0027 & -.016 & -.013 & -.0071 \\
correl.& & & & 0.0081 & -.67 & .24 & -.055 & .029 & -.0048 & .0006 \\
& & & & & 0.0071 & -.62 &.24 & -.067 & .039 & -.0003 \\
& & & & & & 0.0057 & -.60 & .27 &-.028 &.0075 \\
& & & & & & & 0.0045 & -.55 & .19 & .041 \\
& & & & & & & & 0.0032 & -.36 & -.025 \\
& & & & & & & & & 0.0019 & -.0006 \\
& & & & & & & & & & 0.00055 \\
    \hline

  \end{tabular}}
\end{center}
  \label{tab:errmeas}
\end{table}

Similar measurements of the $q^2$ dependence of $\left |f_+(q^2) \right |$
have been obtained by several experiments; recent published results are 
summarized in the following.

Data taken at the $\Upsilon (4S)$ energy and corresponding to an 
integrated
luminosity of $7$ fb$^{-1}$ have been analyzed by CLEO III
\cite{ref:cleo3}. Only three $q^2$ bins have 
been used because of the rather poor resolution obtained on this
variable.

The FOCUS fixed target photo-production experiment has collected 
12840 signal events for the decay chain:
$\Dstarp \rightarrow \Do \pi^+;~ \Do  \rightarrow \Km \mu^+ \nu_{\mu}$
\cite{ref:focus}.
The $q^2$ resolution has a r.m.s. of $0.22~(\GeVc)^2$.

The BELLE $B$-factory experiment has done a measurement by reconstructing
all particles in the event, but the neutrino, which parameters
can be obtained using kinematic constraints. As a result, they achieve a 
very good reconstruction accuracy of about $0.010~(\GeV^2)$ on $q^2$
but at the price of a low overall efficiency. Analyzing 282 fb$^{-1}$
integrated luminosity, they select about 
2500 $D^0 \rightarrow \Km \mu^+ \nu_{\mu}$ and  $D^0 \rightarrow \Km e^+ \nu_{e}$ 
events with low background levels.

In Table \ref{tab:results} results obtained
by the different collaborations have been summarized  when fitting a pole mass 
and a modified pole
mass $q^2$ distributions for the form factor.

\begin{table}[htb]
  \caption {Fitted values for $m_{\rm{pole}}$ and $\alpha_{\rm{pole}}$ models for the form factor.}
\begin{center}
  \begin{tabular}{c|c c c }
    \hline
\hline
Experiment & $m_{\rm{pole}}$ $(\GeVcd)$ & $\alpha_{\rm{pole}}$ & Statistics  \\
\hline
CLEO III \cite{ref:cleo3}  & $1.89 \pm0.05^{+0.04}_{-0.08}$ & $0.36 \pm0.10^{+0.08}_{-0.07}$ & $7$ fb$^{-1}$ \\
\hline
FOCUS  \cite{ref:focus} & $1.93 \pm0.05 \pm 0.03$ & $0.28 \pm0.08 \pm0.07 $ &
13k events \\
\hline
BELLE   \cite{ref:belle}& $1.82 \pm0.04 \pm 0.03$ & $ 0.52\pm0.08 \pm0.06 $ & $282$ fb$^{-1}$  \\
\hline
\babar\  & $1.854 \pm 0.016 \pm 0.020 $ & $ 0.43\pm0.03 \pm0.04 $ & $75$ fb$^{-1}$  \\
    \hline
\hline
  \end{tabular}
\end{center}
\label{tab:results}
\end{table}

Results obtained
in this analysis on the $q^2$ variation of the hadronic
form factor have been compared in Figure \ref{fig:fq2} with
FOCUS \cite{ref:focus} measurements and with the lattice QCD \cite{ref:lqcd}
evaluation. 

The present result is valid in the limit
that in real events the radiative component has the same characteristics
as in the simulation which uses the PHOTOS generator program. 
With the level of accuracy of present
measurements better studies of radiative effects are needed.  

\section{SUMMARY}
\label{sec:Summary}

Fitting the pole mass and the modified pole mass ansatz to the 
measurements, we obtain 
preliminary values for the single parameter that governs their $q^2$
dependence:

\beq
 m_{\rm{pole}} = (1.854 \pm 0.016 \pm 0.020)~\GeVcd \\
 \alpha_{\rm{pole}}= 0.43 \pm0.03 \pm 0.04~~~~~~~~
\nonumber
\eeq
where the first error is statistical and the second systematic.
The effective $m_{\rm{pole}}$ value is rather different from $m_{D_s^{*+}}$
indicating large corrections to naive expectations. In the modified pole model
this can be interpreted as evidence for the contribution from other
vector states of invariant mass higher than the $D_s^{*+}$ mass. 

The value measured for $\alpha_{\rm{pole}}$ agrees, within errors, with the one obtained
from lattice QCD \cite{ref:lqcd}: $\alpha_{\rm{pole}}^{\rm{lattice}}=0.50 \pm 0.04$.
We provide also a preliminary $q^2$ distribution of the form factor, corrected
for effects from reconstruction efficiency and finite resolution
measurements.

\section{ACKNOWLEDGMENTS}
\label{sec:Acknowledgments}


The authors wish to thank C. Bernard and D. Becirevic for a clarification 
on the way to compare the values of the modified pole parameter 
$(\alpha_{\rm{pole}})$ obtained from lattice QCD and from direct measurements.

We are grateful for the 
extraordinary contributions of our \pep2\ colleagues in
achieving the excellent luminosity and machine conditions
that have made this work possible.
The success of this project also relies critically on the 
expertise and dedication of the computing organizations that 
support \babar.
The collaborating institutions wish to thank 
SLAC for its support and the kind hospitality extended to them. 
This work is supported by the
US Department of Energy
and National Science Foundation, the
Natural Sciences and Engineering Research Council (Canada),
Institute of High Energy Physics (China), the
Commissariat \`a l'Energie Atomique and
Institut National de Physique Nucl\'eaire et de Physique des Particules
(France), the
Bundesministerium f\"ur Bildung und Forschung and
Deutsche Forschungsgemeinschaft
(Germany), the
Istituto Nazionale di Fisica Nucleare (Italy),
the Foundation for Fundamental Research on Matter (The Netherlands),
the Research Council of Norway, the
Ministry of Science and Technology of the Russian Federation, and the
Particle Physics and Astronomy Research Council (United Kingdom). 
Individuals have received support from 
the Marie-Curie IEF program (European Union) and
the A. P. Sloan Foundation.

\end{document}